\def\bfseries{\fontseries\bfdefault\selectfont\boldmath}
\def\itshape{\fontshape\itdefault\selectfont\let\mathrm=\mathit}

\RequirePackage{lineno}

\documentclass[preprint,5p,twocolumn,times,sort&compress]{elsarticle}
\pdfoutput=1

\renewcommand{\bibname}{ \vspace{0.3cm} \noindent {\bf References}  $ $}

\usepackage{multirow}
\usepackage{lineno}
\usepackage{subfigure}
\usepackage{graphicx}
\usepackage{epstopdf}
\usepackage{dcolumn} 
\usepackage[retainorgcmds]{IEEEtrantools}
\usepackage{sidecap} 
\usepackage{placeins}
\usepackage{amsmath}
\usepackage{cancel}
\usepackage{tabularx, booktabs}
\newcolumntype{Y}{>{\centering\arraybackslash}X}

\biboptions{numbers,square,comma,sort&compress}
\newcommand{\kpm}{K^{\pm}}

\newcommand{\PD}{p \rightarrow K^{+}\nu}
\newcommand{\eion}{E^{\text{Int}}_{K}}
\newcommand{\edec}{E^{\text{Dec}}_{K}}
\newcommand{\eoth}{E_{\text{Other}}}

\begin{document}

\begin{abstract} 
Large liquid argon TPC detector programs such as LBNE and LAGUNA-LBNO will be able to make measurements 
of the proton lifetime which will outperform Cherenkov detectors in the proton decay channel $\PD$. At the large depths which are proposed for such 
experiments, a non-negligible source of isolated charged kaons may be produced in the showers of cosmogenic muons. We present an estimate of the cosmogenic muon background 
to proton decay in the $\PD$ channel. The simulation of muon transport to a depth of 4 km w.e, is performed in the {\sc MUSIC} framework and the propagation of
muons and secondary particles through to a cylindrical 20~kt LAr target is performed using {\sc Geant4}. 
An exposure time of 100 years is considered, with a rate of $< 0.0012$~events/kt/year at 90\% CL predicted from our simulations.
\end{abstract}

\title{Muon-induced background to proton decay 
in the $\PD$ decay channel with large underground liquid argon TPC detectors.}


\author[sheffield]{J.~Klinger\corref{cor1}}
\ead{j.klinger@sheffield.ac.uk}
\cortext[cor1]{Corresponding author}

\author[sheffield]{V.A.~Kudryavtsev}
\author[sheffield]{M.~Richardson}
\author[sheffield]{N.J.C.~Spooner}

\address[sheffield]{Department of Physics and Astronomy, University of Sheffield, Sheffield S3 7RH, UK}

\maketitle


\section{Introduction}

An attractive framework in which to embed the Standard Model (SM) of particle physics is a Grand Unified Theory (GUT) in which each of
the three independent gauge coupling constants of the SM symmetry group $\text{SU}(3)_\text{C}~\otimes~\text{SU}(2)_\text{L}~\otimes~\text{U}(1)_\text{Y}$
are unified at a high-energy scale, $\Lambda_{\text{GUT}}$.

Any GUT model of hadrons, leptons and gauge interactions would necessarily imply
the violation of baryon-number conservation at the $\Lambda_{\text{GUT}}$-scale~\cite{PhysRevLett.31.661}. If one pursues such a theory, 
the fundamental observables of $\Lambda_{\text{GUT}}$-scale physics would be closely related to the stability of the 
proton. In particular, the lifetime of the proton and the dominant proton decay products could indicate a preference 
towards a specific GUT model and could also provide an insight into physics below the $\Lambda_{\text{GUT}}$-scale.

In the simplest extension to the SM, one can extrapolate the three coupling constants to high energies
such that the different couplings become the same order of magnitude above $10^{15}$~GeV~\cite{Ellis1990441}. One caveat with this procedure is that
the couplings do not unify at a single energy scale~\cite{Amaldi1991447,PhysRevD.36.1385} as one might expect from the 
most simple GUT~\cite{PhysRevLett.32.438}. The observed stability of protons in the decay mode $p \rightarrow \pi^{0} e^{+}$~\cite{PhysRevD.85.112001}
presents another problem for the simplest GUT model. These problems can be avoided in extended GUT models by suppressing or forbidding
this decay mechanism.  Extended GUT models can incorporate supersymmetry~\cite{PhysRevD.26.287}
in which the favoured process for proton decay can be $\PD$~\cite{PhysRevD.38.1479},
whilst allowing the SM couplings to unify at a single energy~\cite{Amaldi1991447}.

Amongst the next generation of high precision proton-decay detectors will be 
large liquid argon (LAr) time projection chambers (TPC) such as LBNE or LAGUNA-LBNO \cite{Adams:2013qkq,1742-6596-375-4-042056}.
These experiments will introduce LAr fiducial volumes of the order of tens of kt at depths of at least $1.5$~km.
LAr TPC imaging will have a higher efficiency to measure the mode $\PD$ compared to 
Cherenkov detectors, and as such they will outperform existing detectors such as Super Kamiokande \cite{PhysRevD.86.012006}.
Given the high efficiency for $\kpm$ particle identification in LAr TPC detectors, the largest background to the $\PD$ signature will be from cosmogenic muons and neutrinos. 

In this paper, we present the results of our simulation of the number of backgrounds to this signal from $\kpm$ mesons
produced in the showers of cosmogenic muons.  We compare our results to the study performed by Bueno {\it et al} in Ref.~\cite{1126-6708-2007-04-041}. 
Compared to the latter study, we perform the first full Monte Carlo simulation of particle production, transport and detection that includes cosmogenic muons and all secondary 
particles\footnote{We define `primary' particles as those which are present at the surface at the Earth, 
and `secondary' particles as those which are produced at any depth below this.}.
We also consider a depth and scale of the detector volume which is more applicable to the LBNE and LAGUNA-LBNO detectors.

\section{Simulation framework}

In our model, a LAr TPC detector is positioned at a depth of 4~km~w.e., which corresponds to the proposed depths of the LBNE and LAGUNA-LBNO detectors. We model
a total mass of 20~kt of LAr, which is close to design specification of the LAGUNA-LBNO detector. 
The detector is modelled as a cylinder of LAr with a diameter of 30~m 
and a height of 20~m. The LAr cylinder in encased within a stainless steel container with a thickness of 5~cm. One can compare this to the study performed by Bueno {\it et al}~\cite{1126-6708-2007-04-041} which reports a prediction of the background rate using a 100~kt volume of LAr at a depth of 3~km~w.e. simulated in the FLUKA~\cite{Ferrari:898301} framework.

The designs of large-scale LAr detectors include up to 2~m of non-instrumented LAr which will separate TPCs from the physical walls. 
This region of non-instrumented LAr defines a `wall' for this study; as no interaction or energy loss occurring in this volume can be seen by the TPC and is unlikely to be caught by a light detector, unless a very sophisticated light detection system is used.
We do not model this region of the detector, and instead treat this boundary as one of the physical walls of the detector.


\begin{figure}[t]
  \centering
  \includegraphics[width=\linewidth]{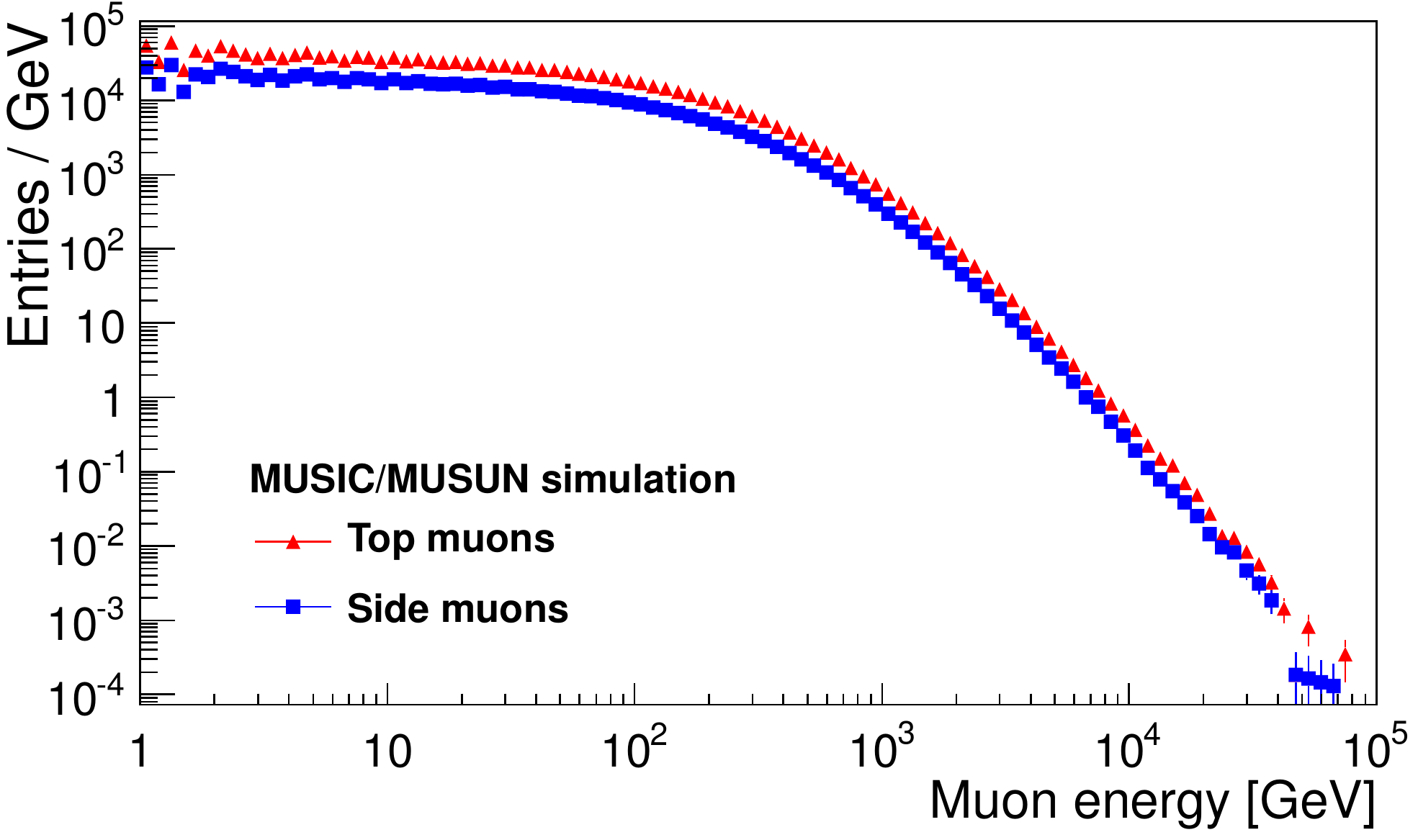}
  \caption{The distribution of muon energies on the top and side surfaces of the cuboid described in the main text, as sampled and simulated by MUSIC/MUSUN~\cite{Kudryavtsev2009339,kud2}. A total of 
  $10^7$~muons are considered in this distribution.}
  \label{fig:muenergy}
\end{figure}

In this paper, the simulation of particle propagation is performed in two stages. In the first stage, only muon transport is considered and the interactions 
of secondary particles are neglected. In the second stage, all particles including secondary particles
are fully simulated. 

In the first stage of the simulation, muons are propagated from the surface of the Earth through a vertical depth of 4~km~w.e. using the MUSIC transport 
code~\cite{Kudryavtsev2009339,kud2}. A simple model of standard rock with proton number $Z=11$, nucleon number $A=22$ and density $\rho = 2.65$~g~cm$^{-3}$ 
is used in the muon transport simulation. The overall muon flux normalisation is provided by the Gaisser parameterisation~\cite{gaisser1990cosmic}, which has been modified
for large zenith angles.
The muons are then sampled on the surface of a cuboid using the MUSUN code~\cite{kud2}. The cuboid has a height of 32~m and horizontal dimensions 
of 40~m~$\times$~40~m, within which the detector cylinder is centred. The energy spectra of a sample of $10^7$ muons which have been fully transported
by MUSIC/MUSUN to the top and sides of the cuboid surface are shown in Figure~\ref{fig:muenergy}. 

The second stage of the simulation is performed using {\sc Geant4.9.6}\footnote{
The {\sc Geant4.9.6} `Shielding' physics list is used, and the muon-nuclear interaction process is additionally switched on.}~\cite{Agostinelli2003250}, in which
all primary and secondary particles are transported from the surface of the cuboid until all surviving particles have exited the cuboid volume. 
In the second stage of the simulation, all particle information and associated energy depositions in the LAr volume are recorded.

\section{Event selection}
\label{sec:evsel}

In this section, we present event selection criteria which maintain high efficiency for selecting $\PD$ signal events whilst rejecting cosmogenic muon-induced backgrounds.

Kaons from free proton decay are expected to have a kinetic energy of 106~MeV. After accounting for the Fermi motion of a proton in a nucleus, the nuclear
binding energy, and subsequent re-scattering of the kaon, the kinetic energy of a kaon emitted from a nucleus of argon is expected to be smeared
across the range 0 to 200~MeV~\cite{Stefan:2008zi}. In a LAr TPC detector, the kinetic energy measurement of charged kaons is further broadened by 
the energy resolution, which we estimate following the procedure presented in Ref.~\cite{1748-0221-6-07-P07011}. In order to retain $> 99\%$ of the signal events we 
require that the energy deposition from a charged kaon due to ionisation and scattering, $\eion < 250$~MeV.

By constraining the total energy of all proton decay products to the proton mass, we require that sum of
$\eion$ and the energy deposition of subsequent decay products, $\edec$, should be less than 1~GeV. Meanwhile energy deposition from other particles not associated with the kaon, its interactions or decay products, $\eoth$, in the same event should be smaller than 50~MeV; otherwise this additional energy deposition would be clearly visible.

We reject charged kaon tracks that originate from within 10~cm of a detector wall to ensure that the candidate signal events from proton decay are not mismatched with the events originating outside of the detector. 
This results in a decrease in efficiency of 2\%. We also require that all tracks are not within 10~cm of the detector walls, to ensure
for proper energy evaluation and particle identification. Events containing such partially-contained tracks are rejected. The requirement for no activity within 10~cm of the detector wall is shown to reject 87\% of events with no primary muons in the fiducial volume.

At the simulated depth of 4~km~w.e., we find that muons pass through the LAr detector at a rate of 0.078 s$^{-1}$, 
so the average time between the two muons crossing the detector is approximately 12.8~s. Assuming that the maximum drift time, 
and hence the duration of the event record, will be about 10~ms, the probability of a muon crossing the detector within any 10~ms time window
will be approximately $10^{-3}$. Rejecting events that contain a muon, where the muon's track length is greater than 20~cm for clear identification, results in the reduction 
of $\PD$ detection efficiency of about 0.1\%.

LAr TPC detectors can identify charged kaons in the range of interest ($< 250$~MeV) via their $\text{d}E/\text{d}x$ in LAr~\cite{1748-0221-6-07-P07011}.
For low energy charged kaons the identification efficiency is assumed to be 100\%, which will lead to a slightly conservative estimate of the background to $\PD$. In order to fully account for the kaon identification efficiency, detailed studies will need to be performed for each specific detector design. 
Although it is assumed that this method can identify charged kaons with a high efficiency, using $\text{d}E/\text{d}x$ provides no separation between $K^+$ and $K^-$ states. The following procedure will provide some degree of separation.
By considering the dominant $\kpm$ decay modes, one would 
 expect to find a $\mu^{\pm}$ in 95\% of $\kpm$ decays, either directly from the $\kpm$ or via a subsequent $\pi^{\pm}$ decay. In these decay 
 modes, $K^+$ and   $K^-$ can be separated in the event of either $\mu^-$~or $\pi^-$~capture in the liquid argon. 
Before considering the muon capture lifetime, the procedure of identifying positive kaons as having a decay chain which features $\mu^{\pm} \rightarrow e^{\pm}$ 
would reject 95\% of negative kaons. After comparing the muon capture lifetime in argon~\cite{1674-1137-38-9-090001} to the muon lifetime~\cite{PhysRevC.35.2212},
the $K^{-}$ rejection factor is reduced to 82\%. In addition to $\kpm$ decay chains featuring $\mu^{\pm} \rightarrow e^{\pm}$, we also permit events featuring $\kpm \rightarrow e^\pm X$ (where $X$ 
is any other set of particles) in order to account for the remaining 5\% of $\kpm$~decays.
 
  \begin{table*}[t]
\footnotesize
\renewcommand{\arraystretch}{1.0} 
\caption{The number of expected events after 100~years of exposure, as a function of sequential selection criteria and the $\kpm$ production mechanism. Events are accepted
into the table if the total energy deposition in the event is less than 2~GeV. Numbers presented in brackets indicate the subtotal number of $K^-$ events. $N$~refers to nucleons in the interaction.}
\centering
\begin{tabularx}{\textwidth}{c *{8}{Y}}
\toprule
$\kpm$ parent & Exactly one $\kpm$ & No muon & No activity near wall & $(\mu^\pm \ \text{or $\kpm$}) \rightarrow e^\pm+X$ &  $\eion < 250$~MeV & $\eion + \edec < 1$~GeV  &  $\eoth < $ 50~MeV \\ 
 \midrule
$\mu^{\pm} + N$ & 255 (43) & 59 (10) & 0 & - & - & - & - \\ 
$\pi^{\pm} + N$ & 134 (20) & 79 (14) & 3 (0) & 3 & 3 & 3 & 0\\ 
$(p / \bar{p}) + N$ & 13 (2) & 7 (1) & 0 & - & - & - & -\\ 
$\gamma$ + N & 8 & 6 & 0 & - & - & - & -\\ 
$\Sigma^-$ & 1 & 1 & 0 & - & - & - & -\\ \hline
$K^0_L$ & 118 (28) & 63 (15) & 31 (2) & 31 (0) & 24 & 24 & 0\\ 
$(n / \bar{n}) + N$ & 11 (1) & 9 (1) & 0 & - & - & - & -\\ 
$K^0_S$ & 12 (2) & 10 (1) & 0 & - & - & - & -\\  \midrule
Total & 552 (96) & 234 (42) & 34 (2) & 34 (0) & 27 & 27 &0\\ \bottomrule
\end{tabularx} 
\label{tab:cutflow}
\end{table*}


To summarise, we propose the following selection criteria for $\PD$ events in large LAr TPC detectors: 

\begin{enumerate}
\item There is exactly one $\kpm$ in the event.
\item There are no muons with track length $> 20$~cm in the detector volume.
\item There is no activity within 10~cm of the detector wall. 
\item The $\kpm$ decay chain includes the decay $\mu^\pm \rightarrow e^\pm$ or $\kpm \rightarrow e^\pm X$, for electrons with $E_e > 5$~MeV.
\item The total energy deposited by the $\kpm$, excluding decay products, satisfies $\eion < 250$~MeV.
\item The total energy deposited by the $\kpm$ and by $\kpm$ decay products, satisfies $(\eion + \edec) < 1$~GeV.
\item The total energy deposited in the rest of the fiducial volume satisfies $\eoth < 50$~MeV.
\end{enumerate}


\section{Results}

\begin{figure}[t]
  \centering
  \includegraphics[width=\linewidth]{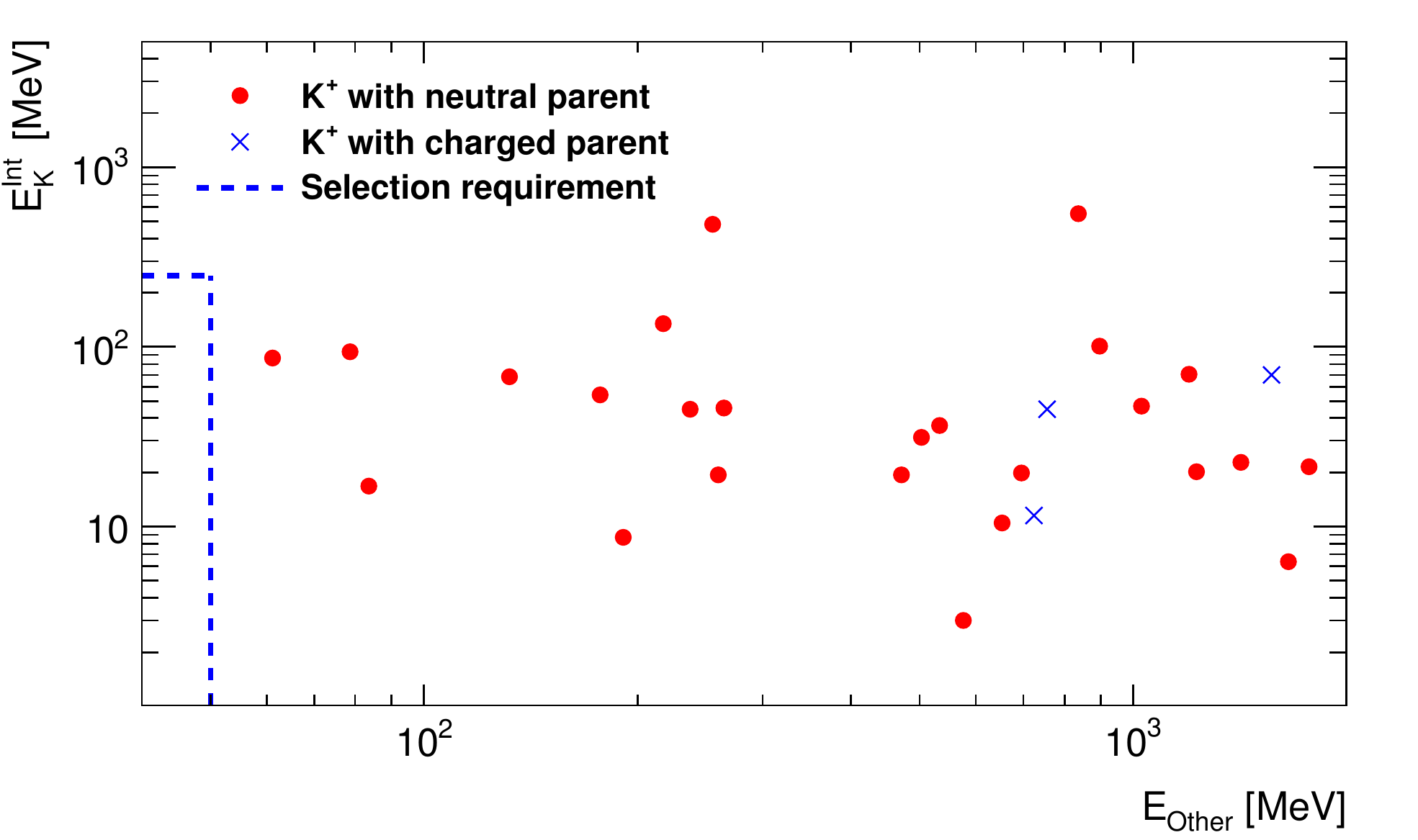}
  \caption{A scatter plot of the total energy deposited by $\kpm$ mesons by ionisation versus the total energy deposition not associated with the kaon in the event. Kaons are selected by applying cuts 1 - 4 as specified in Section~\ref{sec:evsel}. The region retained after the respective selection criteria for the two quantities is bound by the blue-dashed lines. There are no events observed with $\eoth < 50$~MeV. The data represents the statistics after 100~years of exposure.}
  \label{fig:masses}
\end{figure}

In total, an exposure time of 100~years has been simulated. The number of events that are expected to pass the sequential selection criteria described 
in Section~\ref{sec:evsel} is presented in Table~\ref{tab:cutflow}. After applying all cuts, we do not observe
a single muon-induced $\kpm$ event in our simulations. Figure~\ref{fig:masses} 
shows the distribution of $\eion$ and $\eoth$ for events with charged kaons after applying cuts 1 - 4, as specified in Section~\ref{sec:evsel}. The 
region retained after the respective selection criteria for both $\eion$ and $\eoth$ are applied, is bound by the blue-dashed lines. 
It is found that even if the requirement on the kaon kinetic energy is loosened such that  $\eion < 400$~MeV, there will
be no events in the region of interest.

We find that the main source of events 
$\kpm$ events is due to the production of $K_L^0$ outside of the LAr TPC volume. For events passing the selection, the
 $\kpm$ ionisation energy peaks below $200$~MeV, but these events are rejected by requiring that the total energy deposition not associated with the charged kaon 
 is lower than $50$~MeV.

The absence of a single muon-induced $\kpm$ event from our event selection is converted to an 90\% Confidence Level (CL) upper limit of 0.024~background 
events per year in a 20~kt liquid argon detector, or 0.0012~events/kt/year. This result is consistent with earlier estimates~\cite{1126-6708-2007-04-041}. 
The presented event selection reduces the efficiency for $\PD$ selection by less than $2\%$, which is dominated by the requirement for no activity within 10~cm of the detector wall.

\section{Conclusions}

We have presented an estimate of the muon-induced background to proton decay in LAr TPC detectors. 
We consider the background to the process $\PD$, which is favourable in extended GUTs models.

In our simulation, a cylindrical LAr TPC detector is positioned at a depth of 4~km~w.e., which corresponds to the depths of the proposed 
LBNE and LAGUNA-LBNO detectors. We consider a
total mass of 20~kt of LAr, which is close to design specification of the LAGUNA-LBNO detector.

We find that the main source of muon-induced background 
$\kpm$ events is due to muon-induced $K_L^0$ mesons produced outside of fiducial volume. We show that background $\kpm$ events can be rejected by constraining to the kinematics
of the proton decay as well as requiring that the total energy deposition in the event not associated with the $\kpm$ is less than 50~MeV
and that there is no activity within 10~cm of the detector wall. 

After considering an exposure time of 100~years, we set a 90\% CL upper limit on the number of background events as 0.0012~events/kt/year. Furthermore, 
the presented event selection is estimated to reduce the signal efficiency by less than $2\%$, which is dominated by the requirement for no activity within 10~cm of the detector wall.
\FloatBarrier

\section{Acknowledgements}

We thank Andr\'{e} Rubbia for useful discussions, and the LAGUNA-LBNO and LBNE Collaborations for their assistance. This work was partially funded by 
the European Commission
through the FP7 Design Studies with regards to LAGUNA-LBNO (grant number 284518) and the Science and 
Technology Facilities Council consolidated grant (grant number ST/K001337/1). 

\bibname
\bibliography{bib}

\end{document}